\documentclass[aps,prd,superscriptaddress,preprintnumbers,nofootinbib,preprint]{revtex4}

\usepackage{graphicx}
\usepackage{amsmath}

\def\be{\begin{equation}}
\def\ee{\end{equation}}
\def\bea{\begin{eqnarray}}
\def\eea{\end{eqnarray}}

\def\afbt{A_{FB}^t}
\def\afbl{A_{FB}^\ell}
\def\coth{{\text{coth}}}

\begin{document}

\preprint{ANL-HEP-PR-11-74, MSUHEP-111024}

\title{Dynamical Origin of the Correlation between \\the Asymmetries $A_{FB}^t$ and $A_{FB}^{\ell}$}

\author{Edmond L. Berger}
\email{berger@anl.gov}
\affiliation{High Energy Physics Division, Argonne National Laboratory, 
Argonne, IL 60439, U.S.A}

\author{Qing-Hong Cao}
\email{qinghongcao@pku.edu.cn}
\affiliation{Department of Physics and State Key Laboratory of Nuclear Physics
and Technology, Peking University, Beijing 100871, China}

\author{Chuan-Ren~Chen}
\email{crchen@hep.anl.gov}
\affiliation{High Energy Physics Division, Argonne National Laboratory, 
Argonne, IL 60439, U.S.A}

\author{Jiang-Hao Yu}
\email{yujiangh@msu.edu}
\affiliation{Department of Physics and Astronomy, Michigan State University, 
East Lansing, MI 48823, U.S.A}

\author{Hao Zhang}
\email{haozhang@anl.gov}
\affiliation{High Energy Physics Division, Argonne National Laboratory, 
Argonne, IL 60439, U.S.A}
\affiliation{Illinois Institute of Technology, Chicago, Illinois 60616-3793, USA}

\begin{abstract}
A larger than expected forward-backward asymmetry in rapidity is observed in
top quark pairs produced in proton-antiproton collisions at the Tevatron. The asymmetry is
seen in both the top quark distribution $A_{FB}^t$ and in the distribution of charged leptons
$A_{FB}^\ell$ from top quark decay.  In this paper, we study the kinematic and dynamic aspects of the tight 
relationship of the two observables arising from the spin correlation between the charged lepton and the 
top quark with different polarization states.  We also consider two benchmark new physics models, an 
axigluon model and a flavor-changing $W^\prime$ model.  These models could explain the values of 
both $A_{FB}^t$ and $A_{FB}^\ell$.   We emphasize the value of both measurements, and we conclude 
that a model which produces more right-handed than left-handed top quarks is favored by the present data.
\end{abstract}

\maketitle

\section{Introduction}
\label{sec: intro}

The observed forward-backward asymmetry in rapidity $A_{FB}^t$ of top quarks~\cite{Aaltonen:2011kc, Abazov:2011rq} at the Fermilab Tevatron deviates by about two standard deviations ($2\sigma$) from standard model (SM) expectations~\cite{Kuhn:1998jr}.   After corrections for detector acceptance and resolution, $A_{FB}^t$ in the center-of-mass (c.m.) frame is $15.8\pm 7.5 \%$ at  CDF~\cite{Aaltonen:2011kc} and is $19.6\pm6.5 \%$ at D0~\cite{Abazov:2011rq}.    In addition to $A_{FB}^t$, the D0 group also reports a positive forward-backward asymmetry of charged leptons from top quark decays of  $A_{FB}^\ell=(15.2\pm 4.0)\%$ compared with the 
small value $2.1\pm0.1\%$ from simulations of the SM~\cite{Abazov:2011rq}. 
The definition of $A_{FB}^\ell$ is  
\be
A_{FB}^\ell = \frac{n_F -n_B}{n_F + n_B},
\label{eq:def_afbl}
\ee
where $n_F$ ($n_B$) is the number of events with $q_{\ell}y_{\ell} > 0$ $ (q_{\ell}y_{\ell} < 0)$, and $q_{\ell}$ and $y_{\ell}$ are the sign and rapidity respectively of the charged lepton from the semileptonic decay of a top or anti-top quark in the $t{\bar t}$ production.  

In this paper, we investigate the kinematic and dynamic relationship between the two observables 
$A_{FB}^t$ and $A_{FB}^\ell$.   We study quantitatively the influence of the top-quark boost on the kinematics of the 
charged lepton, showing how the distribution of leptons in the laboratory frame is related to the polarization state of the top quark parent.  We show that current data on the ratio of the two asymmetries favor models in which more right-handed than left-handed top quarks are produced.  The fact that $A_{FB}^\ell$, $ A_{FB}^t$, and the ratio $A_{FB}^\ell/A_{FB}^t$ are larger than the SM predictions indicates that the charged lepton strongly prefers to move in the same direction as the top quark from which it originates.  This result can arise if right-handed top quarks~\cite{Krohn:2011tw,Falkowski:2011zr} play a significant role in $A_{FB}^t$ or if a non-standard mechanism produces more highly boosted top quarks at the Tevatron, as we explain below.

Many new physics (NP) models have been proposed to explain the enhancement of $\afbt$, such as flavor-changing $Z^{\prime}$~\cite{Jung:2009jz}, $W^{\prime}$~\cite{wprime} and axigluon $G^{\prime}$~\cite{axi1,axi2,Cao:2010zb} models.%
~\footnote{The next-to-leading order quantum chromodynamics corrections to 
the process of $q\bar{q}\to t\bar{t}$ induced by the flavor-changing $Z^\prime$ and $W^\prime$ are calculated in Ref.~\cite{Xiao:2010hm} and Ref.~\cite{Yan:2011tf}, 
respectively, with the result that the NP prediction at the leading order is reliable. }   
The first two models produce predominantly right-handed top quarks, whereas the axigluon model generates unpolarized top-quarks. 
It is important to validate these models at the Large Hadron Collider (LHC) and/or at the Tevatron.   For example, the heavy flavor-changing $Z^{\prime}$ ($\gtrsim m_t$) model is disfavored because it predicts too much same-sign top quark pair production at the LHC~\cite{Berger:2011ua,Chatrchyan:2011dk}.  In this paper, we focus on how consistently the NP models can describe both $\afbt$ and $\afbl$.   

We begin in Sec.~\ref{sec:kin} with a discussion of the angular distribution of decay leptons, first in the rest frame of the top quark and then after the top quark is boosted in rapidity and transverse momentum.  We pay particular attention to left/right polarization state of the top 
quark because the final distribution of leptons in the laboratory frame, after the top quark is boosted, depends significantly on the top quark's polarization state.  In Sec.~\ref{sec:scan1}, we derive the relationship of the lepton asymmetry $A_{FB}^\ell$ and the top quark asymmetry $A_{FB}^t$ separately for the left- and right-handed polarization states of the top quark.   Different models of new physics produce top quarks with different proportions of left- and right-handed polarization.  We use two such models, an axigluon model and a $W^{\prime}$ model, in Sec.~\ref{sec:scan2} to deduce their different expectations for the ratio of the lepton and top quark asymmetries.  Our conclusions appear in Sec.~\ref{sec:con}.   We emphasize the value of making measurements of both $A_{FB}^t$ and $A_{FB}^\ell$ because their ratio can be related through top quark polarization to the underlying dynamics of top quark production.   

\section{Kinematics}\label{sec:kin}

\begin{figure}
\includegraphics[scale=0.3]{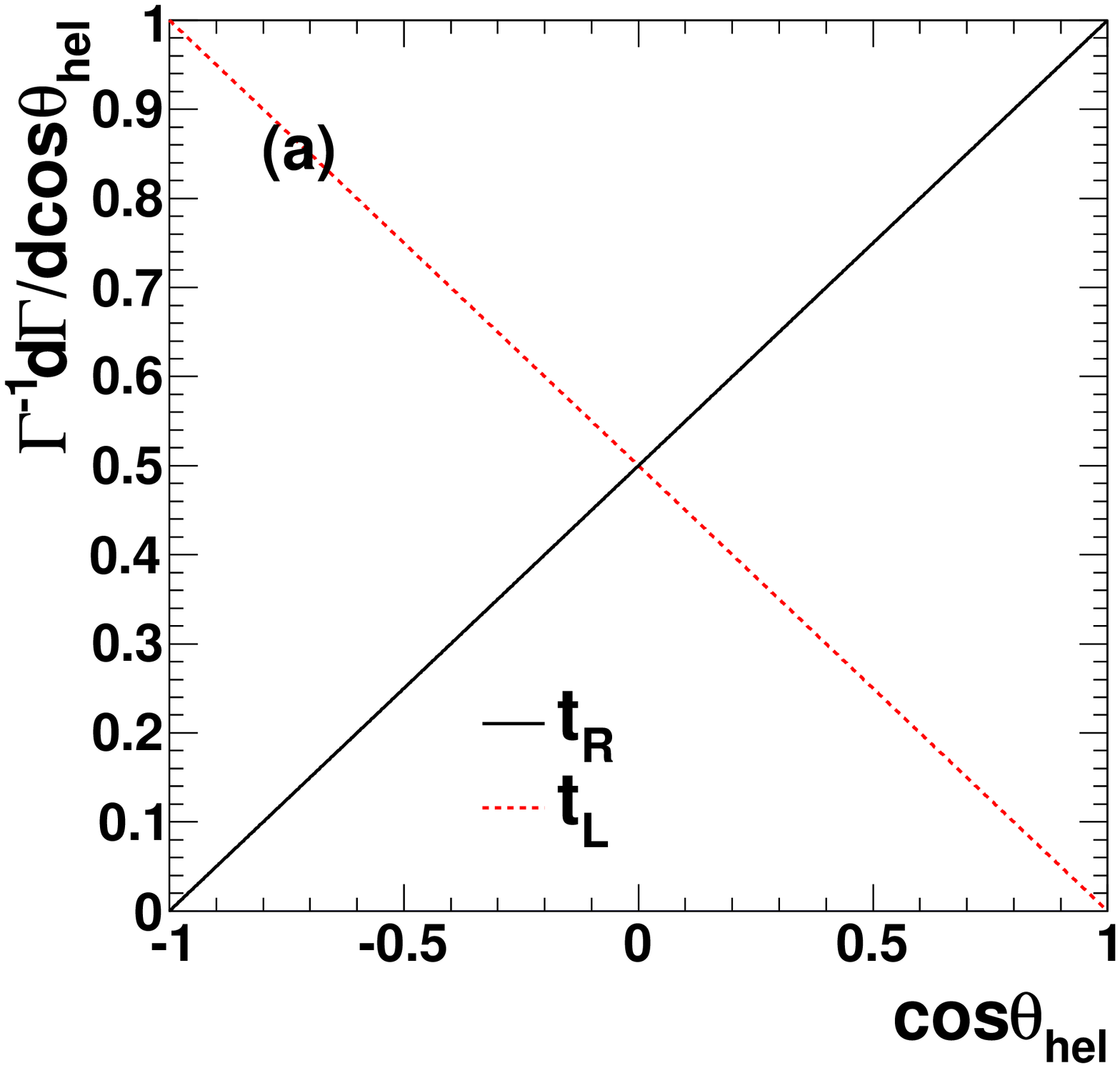}
\includegraphics[scale=0.3]{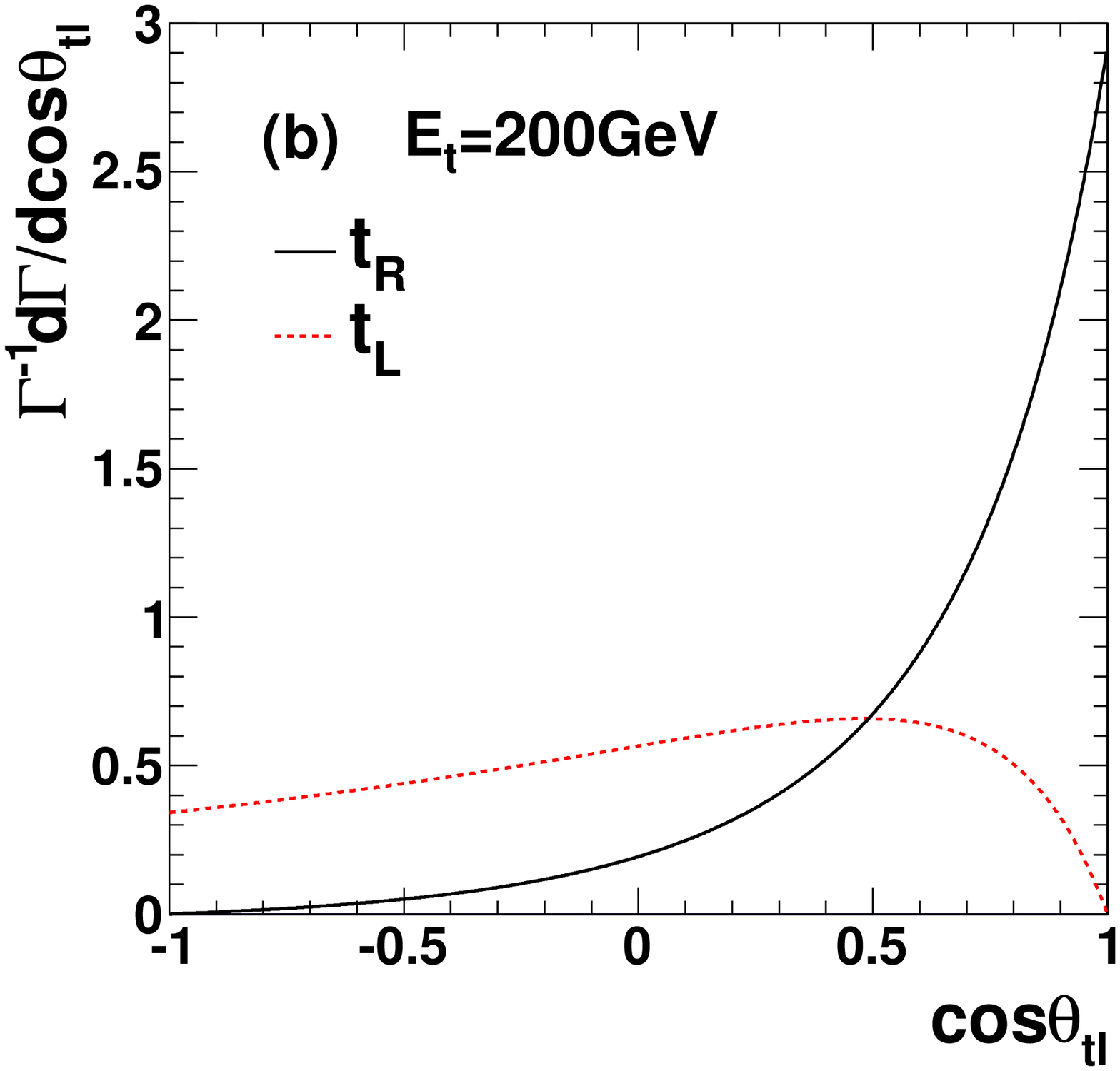}
\caption{(a) $\cos\theta_{\rm hel}$ distribution in the top quark rest frame for both $t_L$ 
and $t_R$. (b) $\cos\theta_{t\ell}$ distribution in the boosted frame for a top quark 
with $E_t=200~{\rm GeV}$.}
\label{fig:leprap}
\end{figure}

The charged lepton in top quark decay is a powerful analyzer of the polarization of the top quark~\cite{Mahlon:1995zn}. 
In the rest frame of a top quark, the distribution in the polar angle  $\theta_{\rm hel}$ of a decay lepton $\ell^+$ is 
\be
\frac{1}{\Gamma}\frac{d\Gamma}{d\cos\theta_{\rm hel}}=\frac{1+\lambda_t\cos\theta_{\rm hel}}{2},
\label{eq:spin}
\ee
where $\lambda_t$ denotes the top quark helicity.  Here, $\lambda_t=+$  is for a right-handed 
top quark ($t_R$) while $\lambda_t=-$ for a left-handed top quark ($t_L$).  The angle is measured 
with resect to the direction of motion of the top quark in the overall center-of-mass system of the 
$t \bar{t}$ production process.   The distributions are shown in Fig.~\ref{fig:leprap}(a).    
The charged lepton from a right-handed top quark decay prefers to move along
the top quark direction of motion, while a lepton from a left-handed top quark moves 
preferentially  against the top quark direction of motion.   In the rest frame of the top quark, 
75\% (25\%) of charged leptons from $t_R$ ($t_L$) decay follow the top quark direction 
of motion, i.e. $\cos\theta_{\rm hel} > 0$.  

Once the top quark is boosted, the angular distribution of the charged lepton relative to the 
direction of motion of the top quark is sensitive to 
the energy of the top quark $E_t$ (or equivalently its velocity $\beta$).  We derive    
\begin{equation}
\frac{d\Gamma}{\Gamma d\cos\theta_{t\ell}}=\frac{1-\beta\cos\theta_{t\ell}+\lambda_t\left(\cos\theta_{t\ell}-\beta\right)}{2\gamma^2\left(1-\beta\cos\theta_{t\ell}\right)^3}, \label{eq:lep_follow_top}
\end{equation}
where $\beta=\sqrt{1-m_t^2/E_t^2}$, $\gamma=E_t/m_t$.
In Fig.~\ref{fig:leprap}(b) we plot the distribution in $\cos\theta_{t\ell}$
of the charged lepton, where the angle $\cos\theta_{t\ell}$ is the angle between the charged 
lepton and its parent top quark in the boosted frame.
As an illustration, fixing the energy of the top quark to $E_t=200~{\rm GeV}$, we find that 
about $60\%$ of $\ell^+$ follow the top quark for a $t_L$, and almost $100\%$ for a $t_R$.

The distribution of charged leptons in the laboratory frame depends on the top quark kinematics, 
including the top quark energy and its rapidity, and the top quark polarization.    The 
probability for finding a positive charged lepton
in the forward region when it originates from a top quark with a velocity $\beta$, rapidity $y_t$, 
and polarization $\lambda_t$  
is defined as 
\be
R_F^{\ell,~\lambda_t}(\beta, y_t)=\frac{N_F^\ell}{N_F^\ell+N_B^\ell},
\ee
where $N_F^\ell$ ($N_B^\ell$) denotes the number of leptons $\ell$ in the 
forward (backward) region in the laboratory.   After lengthy algebra, it can be shown that 
the ratio $R_F^\ell$ is 
\begin{widetext}
\begin{eqnarray}
R_F^{\ell,\lambda_t}(\beta, y_t)=\begin{cases}
\displaystyle \frac{1}{2}+\frac{1}{2\left(1+\gamma^{-2}\coth^2y_t\right)^{1/2}}+\frac{\lambda_t\coth^2y_t}{4\beta\gamma^2\left(1+\gamma^{-2}\coth^2y_t\right)^{3/2}}
&  \displaystyle y_t \in \left[0,~~y_{\rm max}\right]\\
&\\
\displaystyle \frac{1}{2}-\frac{1}{2\left(1+\gamma^{-2}\coth^2y_t\right)^{1/2}}-\frac{\lambda_t\coth^2y_t}{4\beta\gamma^2\left(1+\gamma^{-2}\coth^2y_t\right)^{3/2}},
& \displaystyle y_t \in \left[-y_{\rm max},~0\right]
\end{cases}
\end{eqnarray}
\end{widetext}
where 
\be
y_{\rm max}=\frac{1}{2}\ln\frac{1+\beta}{1-\beta}.
\ee
 
\begin{figure}
\includegraphics[scale=0.3]{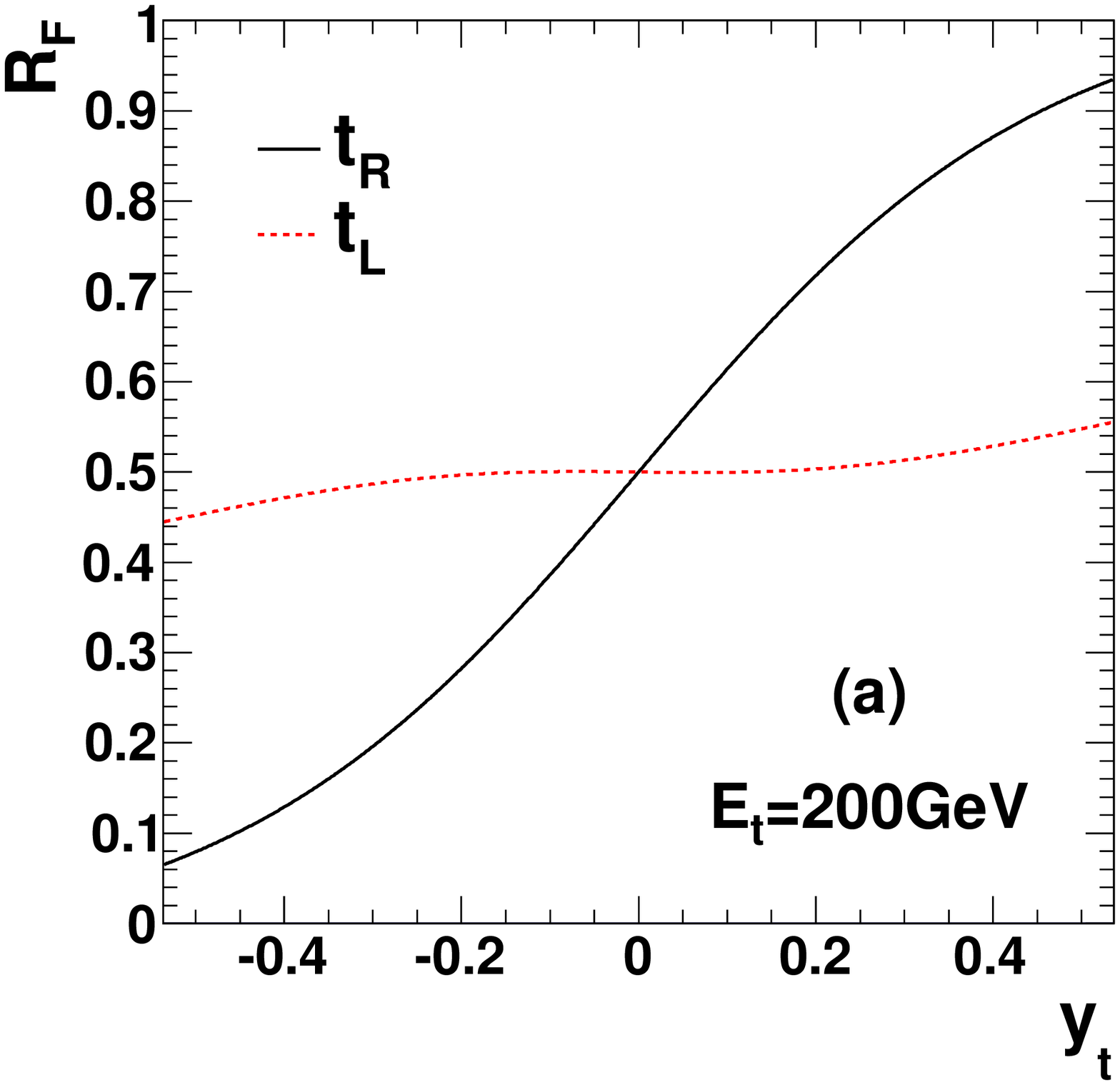}
\includegraphics[scale=0.3]{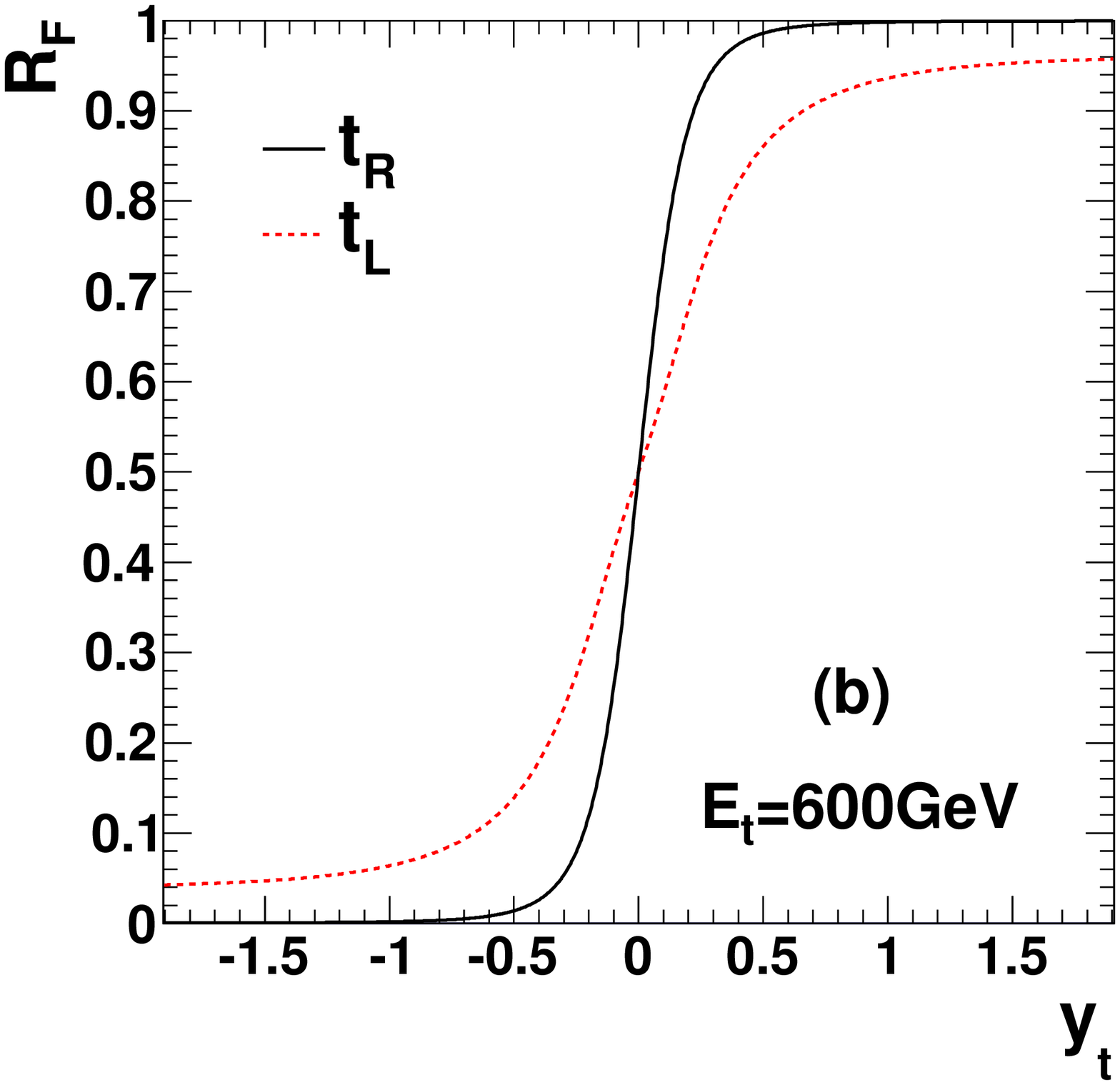}
\caption{
The ratio $R_F$ as a function of $y_t$ for a top quark with fixed energy: 
(a) $E_t = 200~{\rm GeV}$ and (b) $E_t=600~{\rm GeV}$.
\label{fig:lepratio2}}
\end{figure} 
 
To illustrate the effect of the top quark boost, we plot in Fig.~\ref{fig:lepratio2} 
the fraction $R_F$ as a function of $y_t$.  We choose two characteristic top quark energies,
$E_t=200~{\rm GeV}$ and 600~GeV. 
The former energy represents top quarks produced around the threshold region, while the latter 
pertains for highly boosted top quarks. 
Note that $y_{\rm max}=0.53$ for  $E_t=200~{\rm GeV}$. 
When a top quark moves perpendicular to the beam line, i.e. $y_t=0$, there is an equal 
number of leptons in the forward and backward regions, leading  to $R_F=0.5$, independent 
of $E_t$ and the polarization of the top quark.  

For right-handed top quarks $t_R$, $R_F$ increases rapidly with $y_t$ in the region of $y_t > 0$
because most of the leptons move close to the direction of motion of the top quark after being 
boosted to the lab frame; this result is shown by the black solid lines in Fig.~\ref{fig:leprap}.   We 
can also see that when $E_t$ becomes larger, i.e. the top quark is more energetic and the lepton 
is more boosted, $R_F$ rapidly reaches its maximum value $1$. 

On the contrary, in the case of $t_L$'s, the ratio $R_F$ does not vary significantly with $y_t$ owing 
to the anti-boost effect on the charged lepton.  For $E_t = 200$ GeV, the boost causes charged leptons 
to distribute nearly uniformly, and $R_F$ is around $0.5$, as seen in the red-dotted curve in 
Fig.~\ref{fig:lepratio2}(a).   When the energy of $t_L$'s is large enough, the large boost forces most 
of the charged leptons from top quark decays to move along the top quark direction of motion, even 
if they move against the top quark direction of motion in the top quark rest frame.  The boost yields a 
large value $R_F$ in the region of large $y_t$,  as shown by the red-dotted curve in 
Fig.~\ref{fig:lepratio2}(b).  The competing influences leave the $t_L$ curve slightly below the $t_R$ curve.  

\begin{figure}
\includegraphics[scale=0.3]{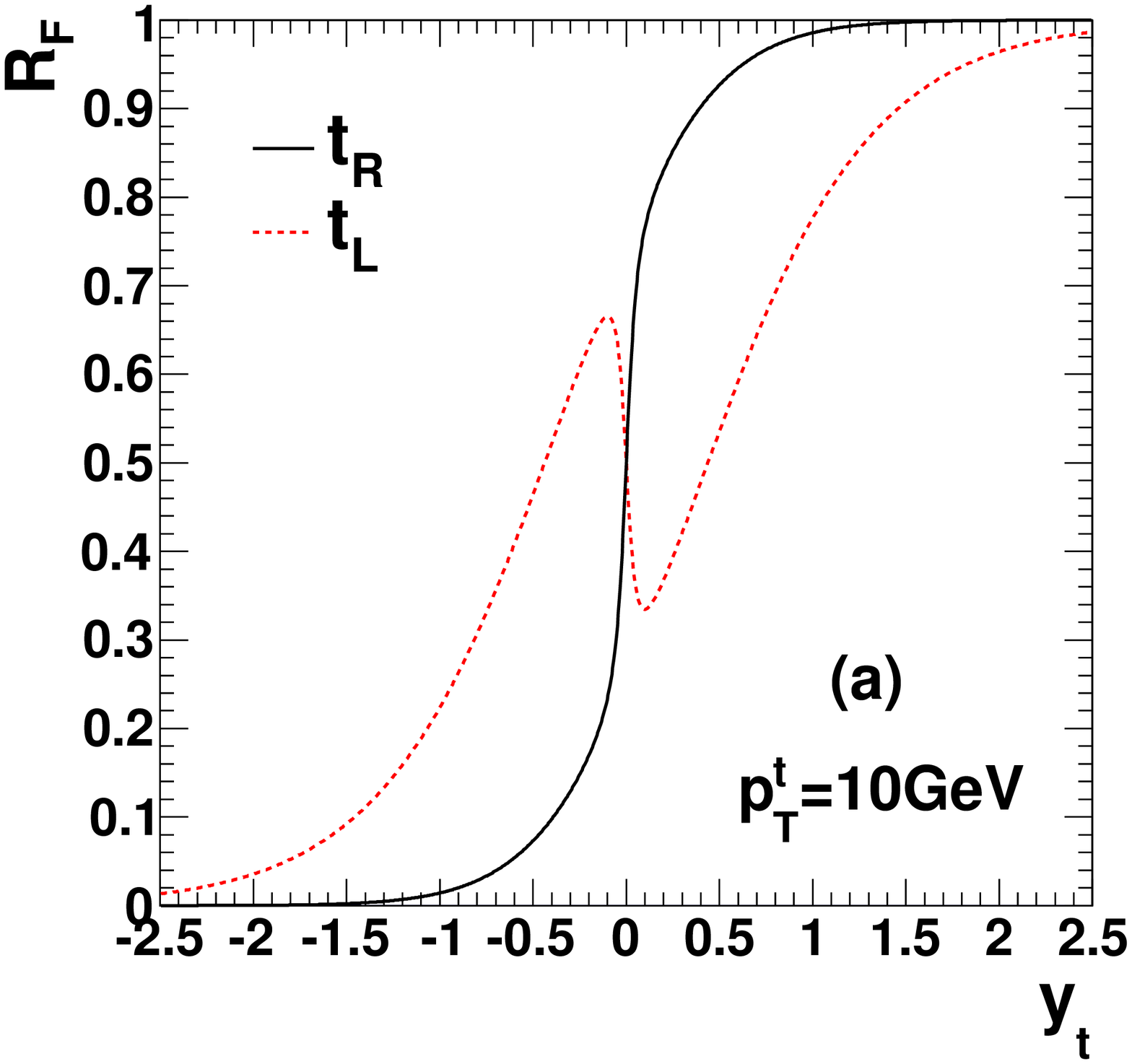}
\includegraphics[scale=0.3]{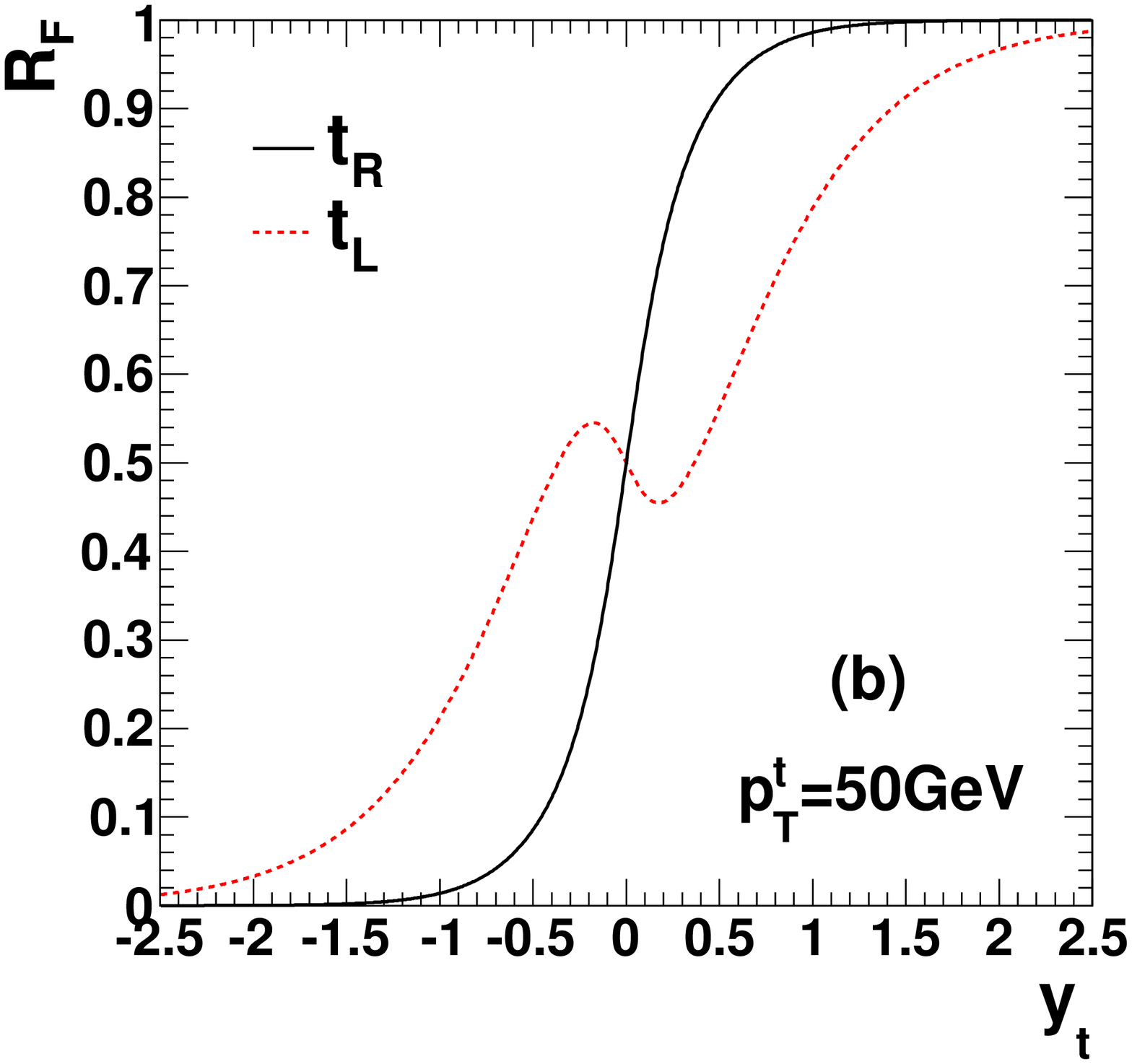}
\includegraphics[scale=0.3]{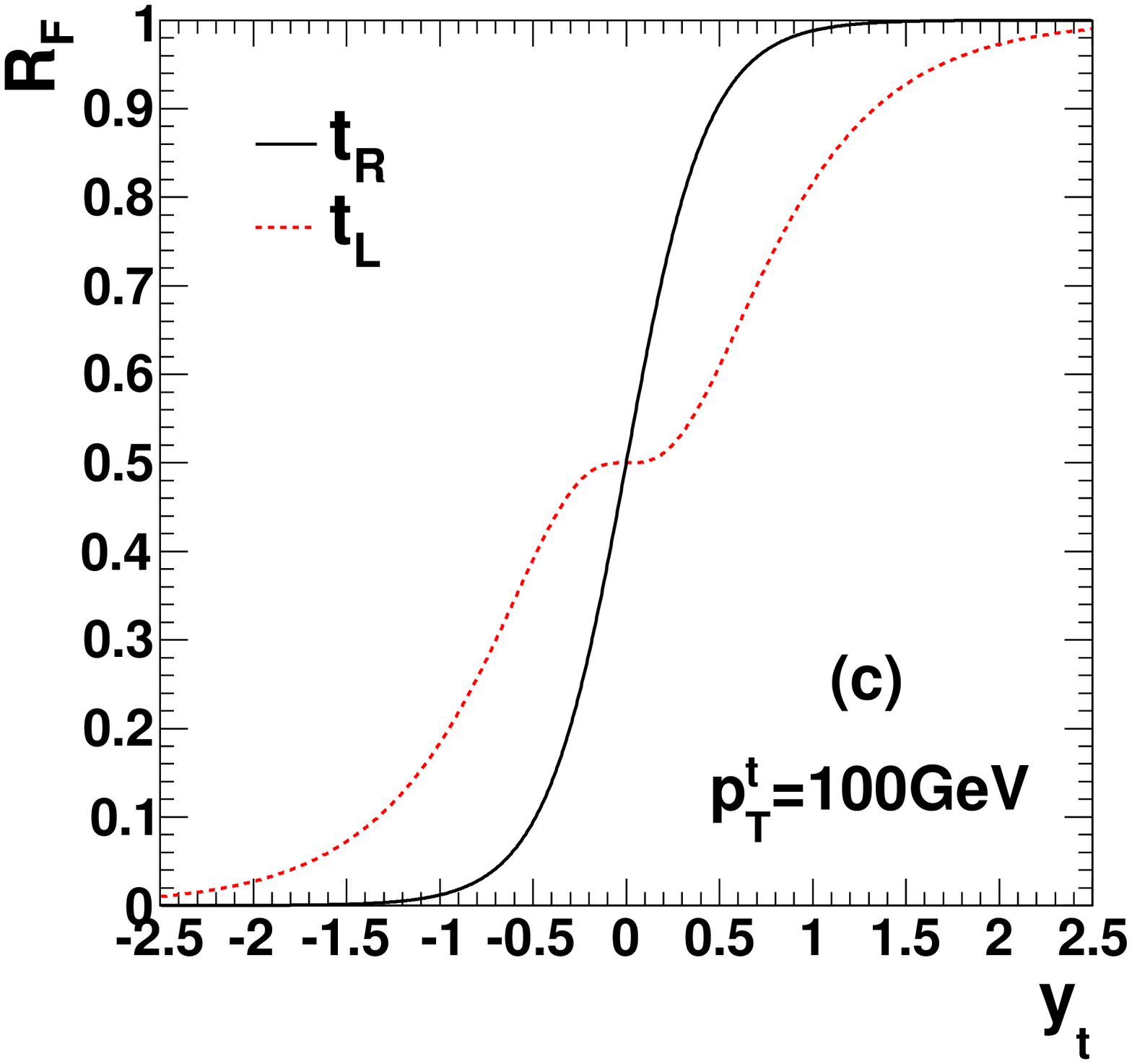}
\includegraphics[scale=0.3]{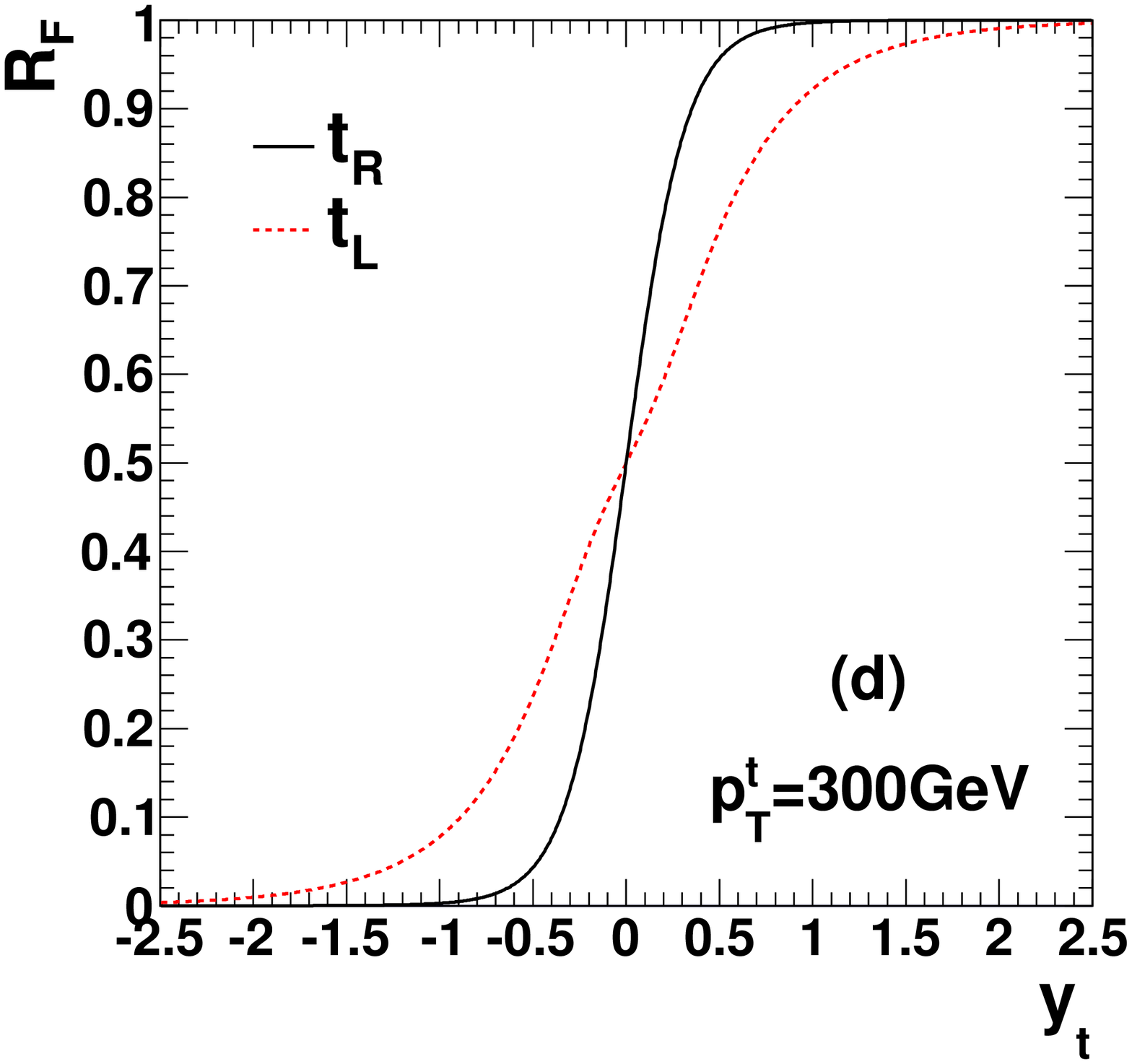}
\caption{The ratio of the charged lepton in the
forward and backward region as a function of the top quark rapidity
for  top quarks with fixed transverse momentum $p_{T} =10, 50, 100, 300~{\rm GeV}$. For a 
fixed $p_{T}= 50~\rm{GeV}$, the figures show that around the region of $y_t \sim 0.2$,  the 
fraction of charged leptons in the forward region is
about $75\%$ for a right-handed top quark while $45\%$ for a left-handed top quark.} 
\label{fig:leprap4}
\end{figure}

In Fig.~\ref{fig:leprap4}, we show how  $R_F$ varies with $p_T^t$ and $y_t$.  The distributions for right-handed 
top quarks $t_R$'s do not vary greatly with $p_T^t$ because most of the charged leptons follow $t_R$.   However, 
the shapes of the curves for left-handed top quarks,  which are the focus in the discussion below,  are very 
different between the low $p_T^t$ and high $p_T^t$ regions, as is seen in the red-dotted lines.  As the top 
quark moves forward, i.e. $y_t > 0$ for fixed $p_T^t$, the boost becomes more significant as the energy of the 
top quark is increased.  Therefore, more leptons are forced to move along the direction of the top quark.  On 
the other hand, some fraction of the decay leptons which are initially in the forward/backward region 
($y_\ell > 0/y_\ell <0$) will then be in the backward/forward region by definition. 
In summary, there are two factors which affect $R_F$: the boost and the rearrangement of the distribution of 
charged leptons in the forward ($y_\ell>0$) and backward ($y_\ell<0$) region. 
The former always increases $R_F$ while the latter may increase or decrease the $R_F$ depending on how 
energetic the top quark is at $y_t =0$.  
Generally speaking, when the boost is not significant (low $p_T^t$ and small $y_t$), $R_F$ decreases when 
$y_t$ increases from $y_t=0$, as we can see in the drop in the red-dotted curves in Fig.~\ref{fig:leprap4}(a) 
and~\ref{fig:leprap4}(b).  When the boost is big enough, $R_F$ always increases with $y_t$.  
The platform-like behavior around $y_t\simeq 0$  in Fig.~\ref{fig:leprap4}(c) arises because the leptons accumulate 
nearly uniformly around the axis of motion of the top quark when $p_T^t =m_t/\sqrt{3}\simeq 100$ GeV.  Therefore the ratio $R_F$
is rather stable as the top quark changes  its direction of motion direction around $y_t =0$. 

\section{$A_{FB}^t$ and $A_{FB}^\ell$}
\label{sec:scan1}

The observed positive top-quark asymmetry $A_{FB}^t$ indicates more top quarks are produced  in 
the forward region than in the backward region of rapidity.
Both $t_R$ and $t_L$ can generate a positive lepton asymmetry $A_{FB}^\ell$.  
However,  as shown in Fig.~\ref{fig:lepratio2}, $t_L$ would need a large boost along the proton beam 
line (i.e. in the large forward rapidity region) to overcome the fact that most of the charged leptons
from its decay move against it in its rest frame.  
A right-handed  top quark $t_R$ can yield a positive $A_{FB}^\ell$ even for top quarks near 
the $t\bar{t}$ threshold region.  Therefore, the large positive top quark and lepton asymmetries 
$A_{FB}^t$ and $A_{FB}^\ell$ observed by the D0 collaboration indicate that
the top quark polarization may be playing a non-trivial role.   In this section we present a general  
analysis of the correlation between $A_{FB}^t$ and $A_{FB}^\ell$, to prepare for a better 
understanding of the numerical results derived from NP models to be shown in Sec.~\ref{sec:scan2}.

The top quark asymmetry $A_{FB}^t$ can be expressed as a sum of contributions from the SM and NP  
as: 
\begin{equation}
A_{FB}^t = A_{FB}^{t,~{\rm  NP}}\times R + A_{FB}^{t,~{\rm SM}}\times (1-R),
\end{equation}
where 
\be
A_{FB}^{t,~{\rm SM}}=\frac{N_F^{\rm SM}-N_B^{\rm SM}}
{N_F^{\rm SM}+N_B^{\rm SM}},~~%
A_{FB}^{t,~{\rm NP}}=\frac{N_F^{\rm NP}-N_B^{\rm NP}}
{N_F^{\rm NP}+N_B^{\rm NP}},~~%
R= \frac{N_{\rm tot}^{\rm NP}}{N_{\rm tot}^{\rm SM}+N_{\rm tot}^{\rm NP}},
\ee
with $N_{F(B)}^{\rm SM}$ and $N_{F(B)}^{\rm NP}$ being the numbers of events in which the 
top quark moves with $y_t>0 (y_t<0)$ in the SM and induced by NP, respectively, and 
$N_{\rm tot}^{\rm SM (NP)}$ is the total number of events predicted in the SM (induced by NP).
The NLO QCD contribution to the production process $q\bar{q}\to t \bar{t}$ could generate 
a value $A_{FB}^{t, {\rm SM}} \sim 5\%$, which is much less than the central value of experimental 
data.  

To somewhat simplify the discussion of the correlation between $A_{FB}^t$ and $A_{FB}^\ell$,    
we assume in this section that $A_{FB}^t$ is generated completely by NP,  but all SM contributions 
(including the NLO QCD effects) are retained in the numerical calculations we present. 

The contributions to $A_{FB}^t$  from different polarizations of top quarks can be separated as: 
\be
A_{FB}^t \approx\left[\rho_{t_L}~A_{FB}^{t_L,~{\rm NP}} +\rho_{t_R}~A_{FB}^{t_R,~\rm NP}\right]\times R,
\ee
where
\be
A_{FB}^{\lambda_t,~\rm NP}=\left[\frac{N_F^{\lambda_t}-N_B^{\lambda_t}}
{N_F^{\lambda_t}+N_B^{\lambda_t}}\right]_{\rm  NP}, \quad
\rho_{\lambda_t} = \frac{N^{\lambda_t,~\rm NP}}
{N_{\rm tot}^{\rm NP}}.
\ee
Here, $A_{FB}^{\lambda_t,~\rm NP}$ denotes the forward-backward asymmetry of the top quark
with polarization $\lambda_t$ generated only by NP, while $\rho_{\lambda_t}$ is
the fraction of  top quarks with polarization $\lambda_t$ in $t\bar{t}$ events induced 
by NP.  One advantage of decomposing $A_{FB}^t$ into different top quark polarizations
is to monitor the chirality of the couplings of NP particles to top quarks. Another advantage is to 
make the connection between $A_{FB}^\ell$ and $A_{FB}^t$ more transparent. 

As discussed in Sec.~\ref{sec:kin} the ratio $R_F^\ell$ depends on the top quark kinematics 
($\beta$, $y_t$ and $\lambda_t$).   To compute the probability for a charged lepton in the 
forward region,  one must convolute the top quark production cross section with $R_F^\ell$ 
on an event-by-event basis, i.e. 
\be
N^{t\bar{t}}\otimes R_F^{\ell,\lambda_t} = \int N^{t\bar{t}}(\beta, y_t, \lambda_t) R_F^{\ell,\lambda_t} (\beta, y_t)~d\Phi,
\ee
where $N^{t\bar{t}}$ labels the $t\bar{t}$ production rate for a top quark with specific kinematics
($\beta$, $y_t$, $\lambda_t$) and $\Phi$ stands for the phase space. 
The lepton asymmetry $A_{FB}^{\ell}$ generated by a top quark with polarization $\lambda_t$ is 
\begin{widetext}
\bea
A_{FB}^{\ell,\lambda_t}\bigg|_{\rm NP} &=& 
\left.
\frac{N_F^{\lambda_t}\otimes R_F^{\ell,\lambda_t}+N_B^{\lambda_t}\otimes R_F^{\ell,\lambda_t}-N_F^{\lambda_t}\otimes R_B^{\ell,\lambda_t}
-N_B^{\lambda_t}\otimes R_B^{\ell,\lambda_t}
}{N_F^{\lambda_t}+N_B^{\lambda_t}}\right|_{\rm NP}
\nonumber \\
&=&\left.\frac{N_F^{\lambda_t}\otimes \left(2 R_F^{\ell,\lambda_t}-1\right)+
N_B^{\lambda_t}\otimes \left(2 R_F^{\ell,\lambda_t}-1\right)}{N_F^{\lambda_t}+N_B^{\lambda_t}}\right|_{\rm NP} \nonumber \\
&=& \left.\frac{(N_F^{\lambda_t}-N_B^{\lambda_t})\otimes
\left(2 R_F^{\ell,\lambda_t}-1\right)}{N_F^{\lambda_t}+N_B^{\lambda_t}}\right|_{\rm NP} .
\label{eq:correlation}
\eea
\end{widetext}
Here, 
\be
R_B^{\ell,\lambda_t}\left(\beta,y_t\right)\equiv \frac{N_B^\ell}{N_F^\ell+N_B^\ell}
=1-R_F^{\ell,\lambda_t}(\beta,y_t),
\ee
and we use the following relation between $R_F^{\ell,\lambda_t}$ and $R_B^{\ell,\lambda_t}$ in our derivation,  
\be
R_B^{\ell,\lambda_t}(y_t) = R_F^{\ell,\lambda_t}\left(-y_t\right). 
\ee
The quantities $N_F^{\lambda_t}$  and $N_B^{\lambda_t}$ in the convolutions in Eq.~(\ref{eq:correlation}) should be understood 
as the distributions $N^{t\bar{t}}(\beta, y_t, \lambda_t)\Theta\left(y_t\right)$ 
and $N^{t\bar{t}}(\beta, y_t, \lambda_t)\Theta\left(-y_t\right)$, respectively,  
where $\Theta\left(x\right)$ is the Heaviside step function.
The quantity $N_F^{\lambda_t}-N_B^{\lambda_t}$ should be understood as 
$\left[N^{t\bar{t}}(\beta, y_t, \lambda_t)-N^{t\bar{t}}(\beta, -y_t, \lambda_t)\right]\Theta\left(y_t\right)$.  
Because $R_F^{\ell,\lambda_t}$ in Eq.~(\ref{eq:correlation}) cannot 
exceed 1, we have $A_{FB}^\ell \lesssim A_{FB}^t$.
When $R_F^{\ell, \lambda_t}$ is close to a constant $\mathcal{R}_C$, 
e.g. $\mathcal{R}_C \sim 1/2$ around the $t\bar{t}$ threshold ($E_t\sim200{\rm GeV}$) for left-handed top quark
or $\mathcal{R}_C\sim 1$ for a highly boosted top quark, 
Eq.~(\ref{eq:correlation}) can be simplified as
\bea
A_{FB}^{\ell, \lambda_t}\bigg|_{\rm NP} &=&\left[\frac{N_F^{\lambda_t}-N_B^{\lambda_t}}
{N_F^{\lambda_t}+N_B^{\lambda_t}}\right]_{\rm NP}
\times   \left(2\mathcal{R}_C-1\right)
=A_{FB}^{\lambda_t,~{\rm NP}} \times  \left(2\mathcal{R}_C-1\right).
\label{eq:correlation2}
\eea
Equation~(\ref{eq:correlation2}) and Fig.~\ref{fig:lepratio2} show that:
\begin{itemize}
\item $A_{FB}^{\ell, t_L} \sim 0$ when the $t\bar{t}$ pair is produced around the threshold region;
\item $A_{FB}^{\ell, t_L} \lesssim A_{FB}^{\ell, t_R} \approx A_{FB}^{t}$ in the large $m_{t\bar{t}}$ region.
\end{itemize}
Although Eq.~(\ref{eq:correlation2}) is approximate, it helps in understanding the NP prediction obtained 
from a complete numerical calculation.

\section{New physics models: axigluon and $W^\prime$}
\label{sec:scan2}

In this section we focus on two models of new physics, an axigluon model~\cite{axi1,axi2,Cao:2010zb} 
and a flavor-changing $W^\prime$ model~\cite{wprime}.   We examine how these NP models can accommodate 
the values of both $A_{FB}^t$ and $A_{FB}^\ell$  measured by the D0 collaboration.  

In the axigluon ($G^\prime$) model we assume for simplicity that the interaction of 
the axigluon to the SM quarks is purely pseudo-vector-like and can be written as
\be
\mathcal{L} = g_s \left(g_l~\bar{q}\gamma^\mu \gamma_5 q 
+ g_h~\bar{Q}\gamma^\mu \gamma_5 Q\right) G^{\prime}_{\mu},
\ee
where $q$ denotes the first two generation quarks in the SM and $Q$ the third generation
quarks.   The coupling $g_s$ is the usual strong coupling strength; $g_l$ and $g_h$ are the 
coupling strength (normalized to the QCD strong coupling $g_s$) of the axigluon to 
the light quark ($q$) and the heavy quark ($Q$), respectively.   

The helicity amplitudes of the processes $q\bar{q} \to g \to t \bar{t}$ and 
$q\bar{q}\to G^\prime \to t\bar{t}$ are written as 
$M_{g}(\lambda_q,\lambda_{\bar q},\lambda_t,\lambda_{\bar t})$, 
and 
$M_{G^\prime}(\lambda_q,\lambda_{\bar q},\lambda_t,\lambda_{\bar t})$, where 
$\lambda_i=+$ represents the right-handed helicity of particle $i$ and $\lambda_i=-$ 
the left-handed helicity.  The total helicity amplitude is
\bea
\mathcal{M}(\lambda_q,\lambda_{\bar q},\lambda_t,\lambda_{\bar t}) =
g_s^2t^A_{ba}t^A_{cd}\biggl[\mathcal{M}_g (\lambda_q,\lambda_{\bar q},\lambda_t,\lambda_{\bar t})
+\frac{\hat{s}~(-g_lg_h)}{\hat{s}-m_{G^\prime}^2 +i m_{G^\prime} \Gamma_{G^\prime}}
\mathcal{M}_{G^\prime} (\lambda_q,\lambda_{\bar q},\lambda_t,\lambda_{\bar t})\biggr],~
\eea
where $t_{ij}^A$ is the generator of the color $SU(3)$ group; $a,~b,~c$ and $d$ are the color indexes of 
$q,~\bar{q},~t$ and $\bar{t}$, 
respectively. The non-vanishing helicity amplitudes are 
\bea
\mathcal{M}_{g}(-+--) &=& -\mathcal{M}_g(+-++)=\sqrt{1-\beta^2}\sin\theta,\nonumber\\
\mathcal{M}_{g}(+---) &=& -\mathcal{M}_g(-+++)=\sqrt{1-\beta^2}\sin\theta,\nonumber\\
\mathcal{M}_{g}(-+-+) &=&~~\mathcal{M}_g(+-+-)=-(1+\cos\theta) ,\nonumber\\
\mathcal{M}_{g}(-++-) &=&~~ \mathcal{M}_g(+--+)=(1-\cos\theta) ,
\label{eq:hel_sm}
\eea
and
\bea
\mathcal{M}_{G^\prime}(+-+-) &=& \mathcal{M}_{G^\prime}(-+-+)= \beta (1+\cos\theta),\nonumber\\
\mathcal{M}_{G^\prime}(-++-) &=& \mathcal{M}_{G^\prime}(+--+)= \beta (1-\cos\theta),
\eea
where $\beta=\sqrt{1-4m_t^2/\hat{s}}$ and $\theta$ is the polar angle of the top quark
in the c.m. frame of the $t\bar{t}$ pair measured relative to the initial state quark. 

The absence of deviation from the SM expectation in the measured $m_{t\bar{t}}$ 
distribution~\cite{Aaltonen:2011kc, Abazov:2011rq}
indicates the axigluon should be heavy and broad.  The axigluon's contribution to $t\bar{t}$ production is therefore 
through interference with the SM channel.  The interference effect becomes largest  in the region of large $m_{t\bar{t}}$,  
i.e. $\beta \sim 1$. Therefore,  the last two equations of Eq.~(\ref{eq:hel_sm}) dominate. 
When $\sqrt{\hat{s}}<m_{G^\prime}$, the denominator of the axigluon propagator is negative, and the 
square of the interference term in the overall amplitude is proportional to 
\be
\left[2g_lg_h(1+\cos\theta)^2-2g_lg_h(1-\cos\theta)^2\right]\frac{\hat{s}}{\hat{s}-m_{G^\prime}^2} .
\ee
The term linear in $\cos\theta$ is $4g_lg_h\hat{s}\cos\theta/(\hat{s}-m_{G^\prime}^2)$.
The product $g_l g_h$ must be negative to obtain a positive 
$A_{FB}$~\cite{axi1,axi2,Cao:2010zb}.  

\begin{figure}
\includegraphics[scale=0.27]{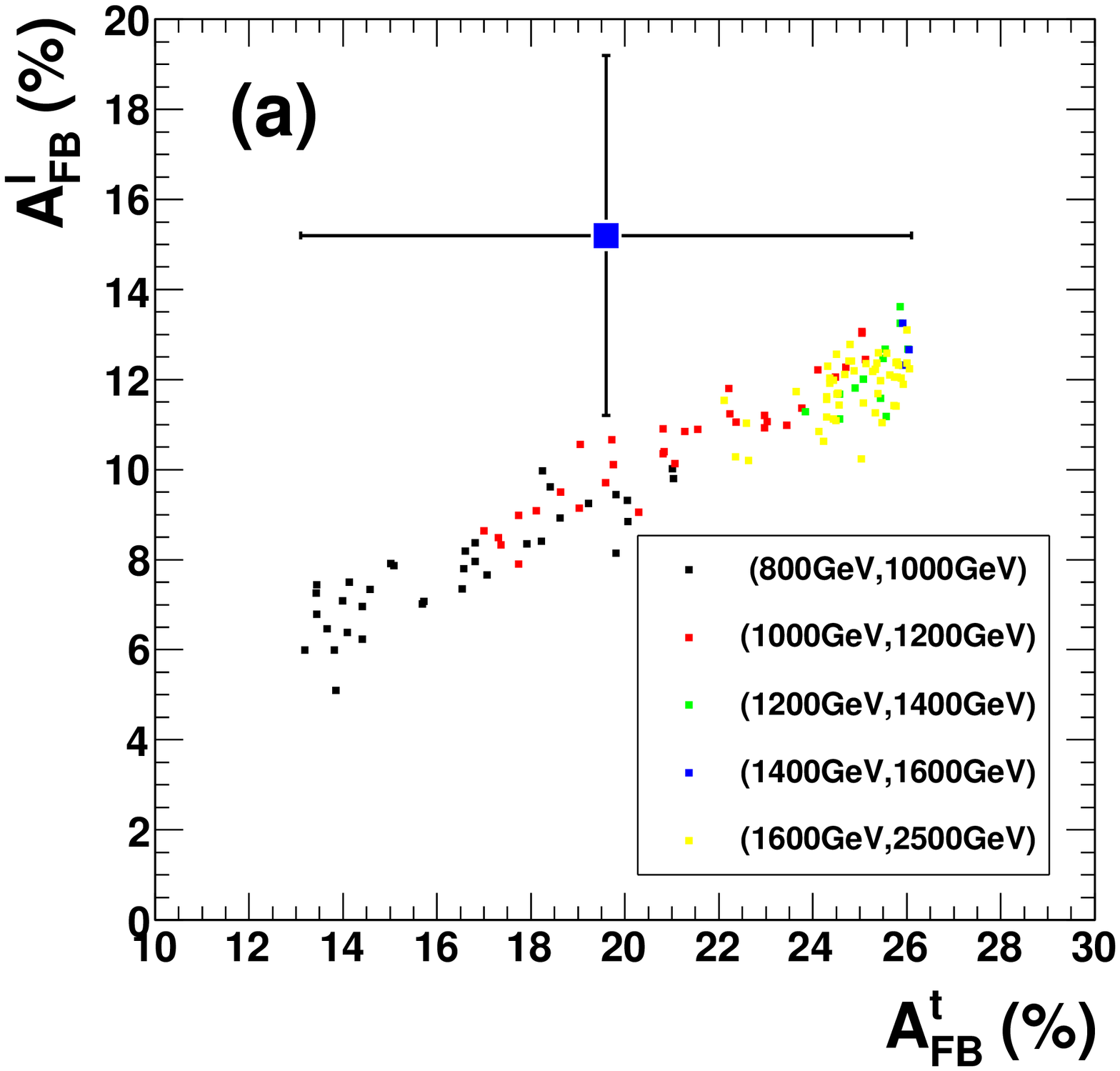}
\includegraphics[scale=0.27]{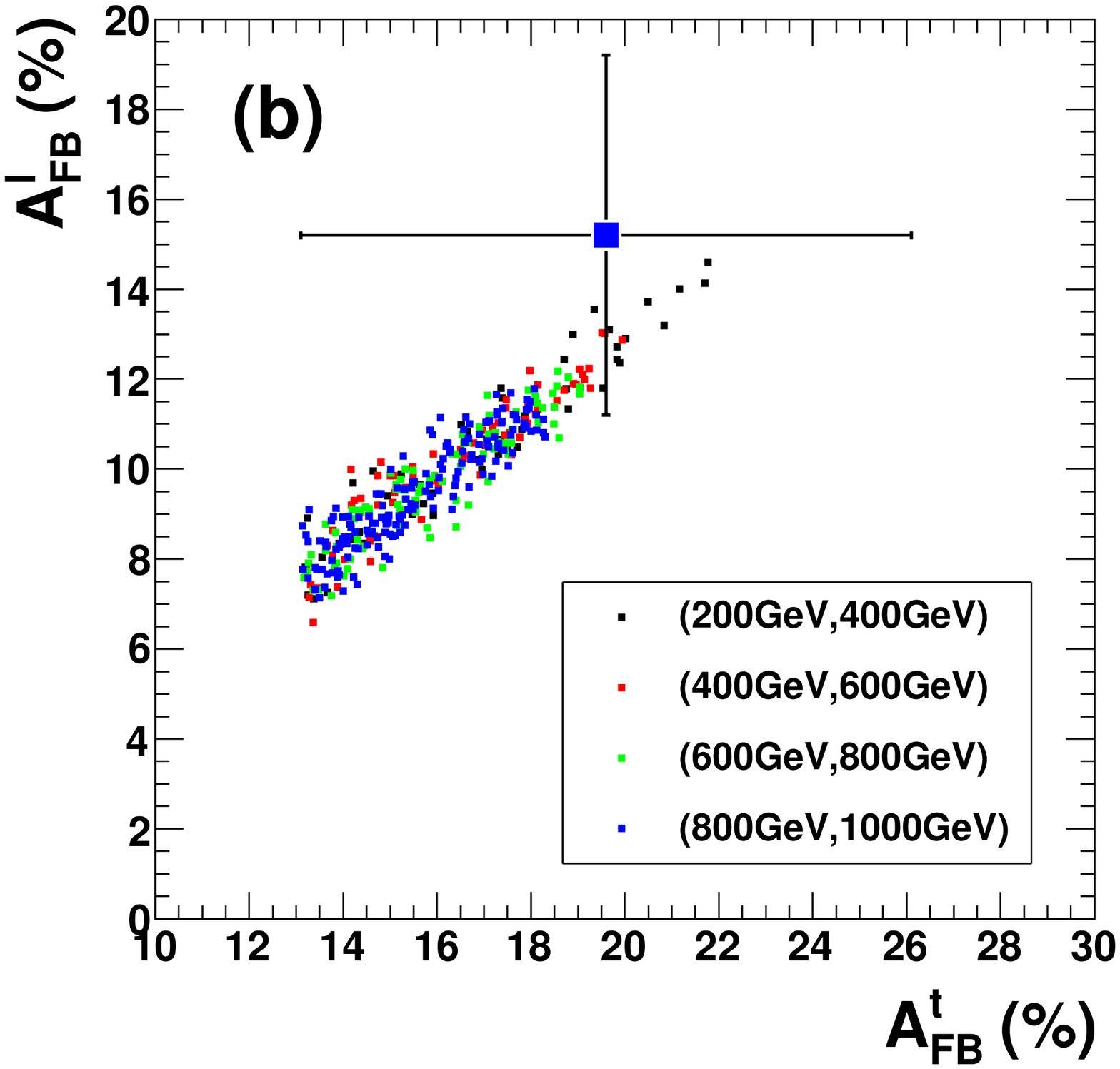}
\includegraphics[scale=0.27]{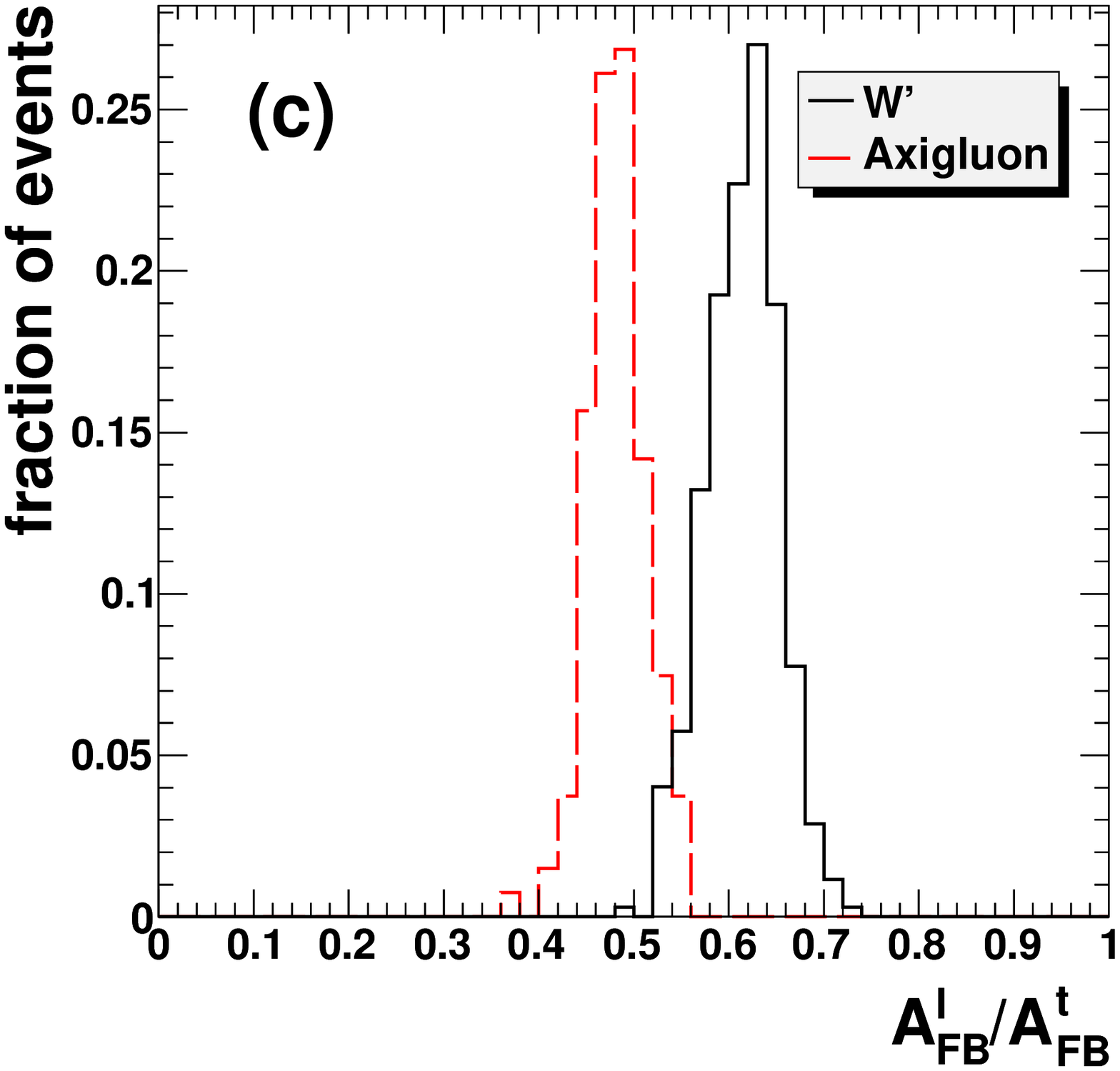}
\caption{Correlation between $A_{FB}^{\ell}$ and $A_{FB}^{t}$ for (a) the axigluon
and (b) the $W^\prime$ models.  The point corresponding to the D0 data is also shown.   
The numbers within the parentheses label the lower and upper limits of the mass of the 
NP  object.   The statistics for the ratio of predicted $\afbl$ to $\afbt$ for the $G^\prime$ 
and $W^\prime$ models are shown in (c).  For comparison, the SM values are 
$A_{FB}^t \sim 5\%$ (off the left side of the plots in (a) and (b), and 
$A_{FB}^\ell \sim 2\%$. }
\label{fig:correlation}
\end{figure}

The top quarks are generated unpolarized owing to the pseudo-vector coupling of 
the axigluon to the SM fermions, and  
 \be
 \rho_{t_L}=\rho_{t_R}=\frac{1}{2},
 ~~~A_{FB}^{t_L,~{\rm NP}}=A_{FB}^{t_R,~{\rm NP}}=\frac{A_{FB}^t}{R}>0.
 \ee
Since the $t \bar{t}$ cross section is greatest near the threshold region where 
$A_{FB}^{\ell, t_L} \sim 0$ and $A_{FB}^{\ell, t_R}\sim A_{FB}^t$, the expression for 
$A_{FB}^\ell$ becomes 
\bea
A_{FB}^{\ell} &\approx& \rho_{t_L} A_{FB}^{t_L,~{\rm NP}} \left( 2 \mathcal{R}_C-1\right)\times R 
+ \rho_{t_R} A_{FB}^{t_R,~{\rm NP}} \left(2\mathcal{R}_C-1\right)\times R 
\nonumber \\
&\sim & \frac{1}{2} A_{FB}^t.
\eea
We plot our axigluon model predictions for $A_{FB}^t$ and $A_{FB}^{\ell}$ 
in Fig.~\ref{fig:correlation}(a).
We first scan the theoretical parameter space ($g_l$, $g_h$ and $m_{G^\prime}$) 
to fit Tevatron data on $A_{FB}^t$ 
and the $t\bar{t}$ total production cross section within  $1~\sigma$. 
These parameters are then used to calculate $A_{FB}^\ell$.  
The figure shows a clear correlation between $\afbt$ and $\afbl$.  The best fit to the 
correlation is
\be
A_{FB}^\ell \simeq 0.47 \times A_{FB}^t + 0.25\%~.
\ee 
To fit both $A_{FB}^t$ and $A_{FB}^\ell$ within $1\sigma$, the mass of the $G^\prime$ 
must be greater than $1$~TeV.  For masses this great,  top quarks from $G^\prime$ decays 
are highly boosted and cause most of  the charged leptons to move along the direction of 
the top quarks.  
We remark here that if the $G^\prime$ is found as a resonance in the $t{\bar t}$ 
mass distribution, the chirality structure of its coupling to $t{\bar t}$ can possibly be determined 
at the LHC~\cite{Berger:2011hn}.
 
A different class of NP models to explain the $t\bar{t}$ forward-backward asymmetry is based 
on $t$-channel kinematics.   Such models involve large flavor-changing interactions. 
A model with a non-universal massive neutral vector boson $Z^\prime$~\cite{Jung:2009jz} is 
one of the possibilities.   However, it is disfavored because it implies a large rate for same-sign 
top quark production at the 7~TeV LHC~\cite{Berger:2011ua}, not supported by 
data~\cite{Chatrchyan:2011dk}.

We consider in this paper a flavor-changing $W^\prime$ which couples an incident $d$-quark to the 
produced $t$-quark~\cite{wprime},
\be
\mathcal{L}=g_2 g_R \bar{d}\gamma^\mu P_R t W^\prime_{\mu} + h.c.~, 
\ee
where $g_2$ is the weak coupling.
In the $W^\prime$ model, in addition to the SM process $q\bar{q} \to g \to t\bar{t}$, 
the $t\bar{t}$ pair can also be produced via a $t$-channel process with
a $W^\prime$ mediator.  Apart from a common factor 
$-i g_2^2 g_R^2 E_t^2/(\hat{t}-m_{W^\prime}^2)$, the helicity amplitude 
$M_{W^\prime}^{t}(\lambda_q,\lambda_{\bar q}, \lambda_{t},\lambda_{\bar t})$ is 
\bea
\mathcal{M}_{W^\prime}^t(+---)&=&-\left[2+r_W^2\right]\sqrt{1-\beta^2} \sin\theta \nonumber \\
\mathcal{M}_{W^\prime}^t(+--+)&=&\left[2(1-\beta)+r_W^2(1+\beta)\right](1-\cos\theta) \nonumber \\
\mathcal{M}_{W^\prime}^t(+-+-)&=&\left[2(1+\beta)+r_W^2(1-\beta)\right](1+\cos\theta) \nonumber \\
\mathcal{M}_{W^\prime}^t(+-++)&=&\left[2+r_W^2\right]\sqrt{1-\beta^2} \sin\theta~,
\eea 
where $r_W=m_t/m_{W^\prime}$.

In the  region $\beta\simeq 1$, the nonzero helicity
amplitudes are
\bea
\mathcal{M}_{W^\prime}^t(+--+)&\sim&2r_W^2(1-\cos\theta), \nonumber \\
\mathcal{M}_{W^\prime}^t(+-+-)&\sim&4(1+\cos\theta)~.
\eea
In order to produce top quarks in the forward region, one needs 
$2r_W^2<4$, which is always true for the region of $W^\prime$ masses (heavier than the top quark) considered in this paper.  
At the Tevatron the $\beta$ distribution of the top quark in $t\bar{t}$ production peaks around $0.6$, and therefore most of the 
top quarks are not significantly boosted.  We can also easily see that more right-handed top quarks are produced compared 
to left-handed ones in the $W^\prime$ model,  $\rho_{t_R} > \rho_{t_L}$.
Since the $t$-channel propagator contributes a minus sign, the total forward-backward asymmetry results from 
a competition between the square of the purely NP term and the interference term of NP with the SM.  The 
former is proportional to $g_R^4$ and the latter to $g_R^2$.    We plot the correlation between $A_{FB}^t$ 
and $A_{FB}^{\ell}$ for the $W^\prime$ model in Fig.~\ref{fig:correlation}(b).   The strong correlation is fit well 
by 
\be
A_{FB}^\ell \simeq 0.75 \times A_{FB}^t - 2.1\%~.
\ee 
Moreover, for a relatively light $W^\prime$ ($\lesssim 600$) GeV, both $\afbt$ and $\afbl$ can be 
consistent with the D0 data within $1~\sigma$. 

For the $G^\prime$ and $W^\prime$ models, Fig.~\ref{fig:correlation}(c) shows the statistics for the ratio of the 
predicted $\afbl$ to $\afbt$,  based on the scattered points in Fig.~\ref{fig:correlation}(a) and ~\ref{fig:correlation}(b).   
The total number of events is normalized to 1.  
The axigluon model peaks near $50\%$ and $W^\prime $ model near $62\%$.   The ratio in the SM is close to 
$40\%$. 
The $W^\prime$ model generates a larger $\afbl$ than the axigluon $G^\prime$ model because it produces 
more right-handed top quarks.  The comparison to the D0 point shown in Figs.~\ref{fig:correlation}(a and b) indicates 
that top quark events with a large proportion of right-handed top quarks are favored.   Constraints on flavor-changing 
currents in the $W^\prime$ model allow only right-handed couplings to the top quark, consistent with the D0 $\afbl$ 
results.   There is no direct evidence of the handedness of the coupling in the massive gluon models.  The D0 result 
could be interpreted as an indirect clue for the chiral couplings of the massive gluon.  Improved statistics would help, 
as well as a measurement of $\afbl$ by the CDF collaboration.  

\section{Conclusion}
\label{sec:con}

The deviation of the top quark forward-backward asymmetry $\afbt$ from its SM prediction may indicate the presence 
of new physics.   Based simply on the large value of $\afbt$, the charged lepton forward-backward asymmetry $\afbl$ 
should also be expected to be larger than the SM expectation.  Indeed, the D0 collaboration reports 
$\afbl = 15.2\%$, about $3\sigma$ away above the SM value.  In this paper, we study the kinematic and dynamic aspects 
of the relationship between the asymmetries $\afbt$ and $\afbl$ based on the spin correlation between charged leptons 
and the top quark with different polarization states.   Owing to the spin correlation in top quark decay, 
$\afbl$ and $\afbt$ are strongly positively correlated for {\em right-handed} top quarks.   However, for {\em left-handed} 
top quarks, the nature of the correlation depends on how boosted the top quark is.   For large enough top quark energy, left-handed 
top quarks will also generate a large charged-lepton asymmetry, similar to that for right-handed quarks.  However, if the 
top quark is not boosted ($E_t \lesssim 200$ GeV), $\afbl$ from left-handed top quarks will be less than $\afbt/2$ for a 
positive $\afbt$.  Since most of the $t\bar{t}$ events are  produced in the threshold region, one may use the large positive 
values of $\afbt$ and $\afbl$ measured at D0 to conclude that production of left-handed top quarks is disfavored.   
Confirmation of the D0 result and greater statistics are desirable.   There is great value in making measurements of 
both $A_{FB}^t$ and $A_{FB}^\ell$ because their correlation can be related through top quark polarization to the underlying 
dynamics of top quark production.   

We focus on two benchmark NP models, 
an axigluon ($G^\prime$) model which produces unpolarized top quarks, i.e. an equal number of right-handed and 
left-handed top quarks, and a flavor-changing $W^\prime$ model which produced dominantly right-handed top 
quarks.   To determine free parameters, we require that these new physics models fit $\afbt$ as well 
as the $t{\bar t}$ total cross section at the Tevatron at $1\sigma$ level of accuracy.  As we show, there is a  
strong correlation between $\afbt$ and $\afbl$ in both models.   The best fit to the relationship is 
$A_{FB}^\ell \simeq 0.47 \times A_{FB}^t + 0.25\%$ and  $A_{FB}^\ell \simeq 0.75 \times A_{FB}^t - 2.1\%$, for the axigluon model 
and the $W^\prime$ model, respectively, both within $2\sigma$ of the D0 result.   To generate $\afbl$ satisfying the data 
to better than $1\sigma$ accuracy, a heavy $G^\prime$ (heavier than about $1$ TeV) is preferred, and a light $W^\prime$ 
(lighter than $600$ GeV) is favored. 

We do not address the LHC case in this paper but may do so at a later time.  Owing to the lack of definition of a forward direction  
in a $p p$ collision, it is less straightforward to measure the two observables we discuss here. 

\begin{acknowledgments}~The work of E.L.B., C.R.C. and H.Z. is supported in part by the U.S.
DOE under Grants No.~DE-AC02-06CH11357. H.Z. is also supported by DOE under the Grant No. DE-FG02-94ER40840.    
The work of J.H.Y. is supported in part by the U.S. National Science Foundation 
under Grand No. PHY-0855561.
\end{acknowledgments}

\end{document}